\documentclass[a4paper,twocolumn,deluxetable]{aa}
\usepackage{deluxetable}
\usepackage{graphicx} 
\usepackage{txfonts} 
\usepackage{epsf} 
\def\sec{\ifmmode {}^{\prime\prime}\else ${}^{\prime\prime}$\fi~}

\def\magdot{\ifmmode {}^{\rm m}\!\!\!.\, \else ${}^{\rm m}\!\!\!.\,$\fi} 
\def\daydot{\ifmmode {}^{\rm d}\!\!\!.\, \else ${}^{\rm d}\!\!\!.\,$\fi} 
\def\asec{\ifmmode ^{\prime\prime}\else$^{\prime\prime}$\fi} 
 
\begin{document} 
 
\title{The Nainital-Cape Survey -- II:\\ Report for pulsation in five chemically peculiar A-type stars and presentation of 140 null results}

\author{Santosh Joshi \inst{1,2}, D. L. Mary \inst{2,3}, Peter Martinez \inst{4}, D. W. Kurtz \inst{5}, \\
V. Girish\inst{6}, S. Seetha \inst{7}, Ram Sagar \inst{2}, B. N. Ashoka \inst{7}}

\offprints{Santosh Joshi, \email{santosh@iucaa.ernet.in}}

\authorrunning{Santosh Joshi et al.} 

\institute{ 
Inter-University Centre for Astronomy and Astrophysics (IUCAA), Post Bag 4, 
Ganeshkhind, Pune 411007, India\\ 
\email{santosh@iucaa.ernet.in}
\and  
Aryabhatta Research Institute of Observational Sciences (ARIES), Manora Peak, 
Nainital-263129, India\\
\email{santosh@aries.ernet.in} 
\and 
Astronomisches Rechen-Institut am Zentrum fuer Astronomie, Moenchhofstrasse 12-14, D-69120, Heidelberg, Germany\\
\email{dmary@ari.uni-heidelberg.de}
\and 
South African Astronomical Observatory (SAAO), PO Box 9, Observatory 7935, South 
Africa\\
\email{peter@saao.ac.za} 
\and 
Centre for Astrophysics, University of Central Lancashire, Preston PR1 2HE, UK \\ 
\email{dwkurtz@uclan.ac.uk} 
\and 
Tata Institute of Fundamental Research (TIFR), Homi Bhabha Road, Mumbai-400 005, India \\ 
\email{giri@tifr.res.in} 
\and 
ISRO Satellite Center, Airport Road, Bangalore-560 017, India \\ 
\email{seetha@isac.ernet.in}}
\date{}

\abstract
{}
{To search photometric variability in
chemically peculiar A type stars in the northern hemisphere.}
{High-speed photometric observations of Ap and Am star candidates have been
carried out from ARIES (Manora Peak, Nainital) using a three-channel fast
photometer attached to the ARIES 104-cm Sampurnanand telescope.}
{This paper presents three new variables: HD\,113878, HD\,118660 and HD\,207561. During the time span of the survey (1999 December to 2004 January) pulsations of the $\delta$\,Sct type were also
found for the two evolved Am stars HD\,102480 and HD\,98851, as reported in Joshi et al. (2002, 2003). Additionally, we present 140 null results of the survey for this time span.}
{The star HD\,113878 pulsates with a period of $2.31$\,hr, which is
typical of $\delta$\,Sct stars. HD\,118660 exhibits multi-periodic
variability with a prominent period of nearly $1$\,hr. These periods need to
be investigated and make HD\,118660 a particularly interesting target for
further observations. For HD\,207561, a star classified as Am, a probable
pulsation with a period of 6\,min was found in the light curves obtained on
two consecutive nights. 
Both HD\,102480 and HD\,98851 exhibit unusual alternating high and low amplitude maxima, with a period
ratio of 2:1.  The analysis of the null results confirms  the photometric quality of the Nainial site.}


\titlerunning{The Nainital-Cape Survey -- II} 
\maketitle 

\section{Introduction} 
The Nainital-Cape Survey is a collaborative survey program to search for
pulsational variability in chemically peculiar A-type stars in the northern
hemisphere. It was initiated in 1997 between the Aryabhatta Research
Institute of Observational Sciences (ARIES -- formerly State Observatory),
Nainital, India, and the South African Astronomical Observatory (SAAO),
Cape Town, South Africa (Seetha et al. \cite{seetha01}). Details of the
facilities at ARIES have been published by Sagar and Mary (2005). This paper, in
which we give the current status of the survey, is the second of a series.
The methods and first results of the survey were published by Martinez et
al. (\cite{martinez01}; hereafter, Paper I).

The class of A stars contains diverse stars that range from radiative photospheres at A0 to mainly convective photospheres by A9. The physics of these stars still challenges our knowledge in many respects including convection, the effects of internal rotation (Yildiz \cite{yildiz03}; Reiners \& Royer \cite{reiners04}), the coupling of rotation with magnetic field (Arlt \cite{arlt04}) and with chemical mixing (Noels et al. \cite{noels04}), the origin of the magnetic field, or the mechanisms responsible for the pulsations observed in some of these stars (Kurtz \cite{kurtz00}; Balmforth et al. \cite{balmforth01}; Cunha \cite{cunha02}; Saio \cite{saio05}; Cunha \cite{cunha05}).

The diversity in the nature of A stars makes it very difficult to draw a clear line between ``normal'' and ``peculiar'' stars. By a normal A-type star, it is generally understood that at classification dispersion the star shows none of the anomalies characteristic of other classes of stars; that when subject to a local thermodynamic equilibrium analysis, it appears to have a composition like the Sun's; and that it exhibits no variability (Wolff \cite{wolff83}). The problem of the resolution required to distinguish normal from peculiar stars is an old, though somewhat recurrent one. Those stars which present peculiar elemental abundances, the chemically peculiar (CP) stars, are mostly
A-type stars, but range from early-B to early-F.

Their abundance anomalies are generally detected by the presence of abnormally strong and/or weak absorption lines of certain elements in their optical spectra. The peculiarity in these stars is interpreted as atmospheric under-abundance and over-abundance of different chemical elements, and is explained quite successfully by the {\it{diffusion process}} (Michaud \cite{michaud70}; Michaud \& Proffitt \cite{michaud93}; Vauclair \cite{vauclair04}). As outlined by Dworetsky (\cite{dworetsky04}), the observed spectra are connected to changes in the deep interior of the stars, and diffusion cannot be regarded as only a surface phenomenon. 

In the regime of CP stars, some Ap (A-peculiar) and Am (A-metallic line) stars in the spectral range A-F, lying within or near to the $\delta$\,Sct instability strip, show p-mode photometric variability in the period range of 5.65\,min to 8\,hr. Those Ap stars which exhibit non-radial, low-degree, high-overtone acoustic pulsations with observed periods between $5.65 - 21$\,min are known as rapidly oscillating Ap (roAp) stars (Kurtz \cite{kurtz82}, \cite{kurtz90}; Kurtz \& Martinez \cite{kurtzm00}). Those variables with variability periods of $18.12$\,min (HD 34282, Amado et al. \cite{amado04}) to about $8$\,hr are known as $\delta$\,Sct stars (see Breger \cite{breger00} for a review of these stars). The majority of $\delta$\,Sct stars are non-radial, low-order, low-degree, pressure ($p$) mode
pulsators. Both roAp and $\delta$\,Sct-like CP stars are of interest for the present survey.

A total of 35 members of the class of roAp stars have been discovered so far. One of them, HD\,12098, was discovered in the framework of the
Nainital-Cape survey; see Paper I and Girish et al. \cite{girish01}. These stars exhibit strong global magnetic fields. Recent results show that the distribution of the number of stars versus $\langle H \rangle$ is close to a negative exponential with very few stars having more intense fields than 8\,kG (Dworetsky \cite{dworetsky04}). The second largest magnetic field in an Ap star (24.5\,kG) was discovered recently in HD\,154708 (Hubrig et al. \cite{hubrig05}); this star is also a newly-discovered roAp star (Kurtz et al., in preparation). The predominant structure of the field is dipolar, although deviations from the dipolar case have been discovered by Leroy et al. (\cite{leroy95}) and modelled by Bagnulo et al. (\cite{bagnulo99}). 

The Ap stars exhibit chemical abundance anomalies, particularly overabundances of Rare-Earth elements. Since the discovery that chemical peculiarity and pulsation can coexist in some magnetic CP stars (see Kurtz 1982), many attempts have been made to understand the geometry and driving mechanisms of the corresponding pulsations. The pulsation axis is thought to be aligned with the magnetic axis, and is consequently oblique to the rotation axis: the {\it oblique pulsator model}. See Kurtz 1982, 2000; Takata \& Shibahashi \cite{takata95}; Saio \& Gautschy 2004; and Saio 2005 for discussion of the oblique pulsator model, and also see Bigot \& Dziembowski (\cite{bigot02}) who find that the axis of pulsation is not necessarily aligned with the magnetic axis in the case of relatively weak magnetic fields, $H_s \le 1$\,kG, when centrifugal effects from rotation may dominate. 

In order to explain the driving mechanism of the short-period, high-overtone
modes of roAp stars, many works have turned to the $\kappa$ mechanism. This
mechanism acts for $\delta$\,Sct stars in the He\,\textsc{ii} ionization
zone (Chevalier \cite{chevalier71}) and leads to long periods corresponding
to fundamental/low-overtone modes; as such, it is not able to explain
directly the short-period pulsations of roAp stars. Several works have
investigated the respective influence of the magnetic field on Helium settling
and on convection efficacy (Dolez \& Gough \cite{dolez82}), stellar wind
likely to activate the 
$\kappa$ mechanism in the He\,\textsc{i} ionization zone (Dolez et al.
\cite{dolez88}), the possible role played by the ionisation of
Si\,\textsc{iv} (Matthews \cite{matthews88}), the $\kappa$ mechanism acting
in the H\,\textsc{i} ionization zone (Dziembowski \& Goode
\cite{dziembowsky96}), or mechanisms based on the possible existence of
chromospheres (Gautschy, Saio \& Harzenmoser \cite{gautschy98}). Apart from
the $\kappa$ mechanism, some works have investigated other possible causes
for driving the pulsations, such as the Lorentz force (Dziembowski \& Goode
\cite{dziembowsky85}), overstable convective modes (Shibahashi
\cite{shibahashi83}, Cox \cite{cox84}), or stochastic excitation (Houdek et
al. \cite{houdek99}). The current understanding is that the $\kappa$ mechanism
acting in the region of the first ionization of Hydrogen is probably the
driving mechanism of pulsations in roAp stars, acknowledging the facts that
in this case oscillations are aligned with the magnetic axis, that this
model leads to frequencies as high as those observed in roAp stars, and that
the diffusion-induced Helium gradient of this model may lead to particular
asteroseismic signatures\footnote{i.e. modulation of frequencies due to
partial reflection of the sound wave where the He gradient occurs, e.g. in
HD\,60435 (Vauclair \& Th\'eado \cite{vauclairetal04}).} as observable in
some roAp stars. Using these ideas, Balmforth et al. (\cite{balmforth01})
have proposed a model for the excitation mechanism in roAp stars where the
effect of the magnetic field is to freeze the convection at the poles. On
the basis of this model, Cunha (\cite{cunha02}) calculated the theoretical
boundaries of the instability strip in the Hertzsprung-Russell (H-R)
diagram, and provided many theoretical insights on the pulsations of these
stars. Saio (2005) calculates that high-overtone p modes are excited by the
$\kappa$ mechanism for H\,\textsc{i}, and that low-overtone p modes –- those
typical of $\delta$\,Sct stars -– are not excited in the presence of a
magnetic field. These results are consistent with the short periods of the
roAp stars and the lack of $\delta$\,Sct pulsations found in {\it confirmed}
magnetic stars.

 So far, many Ap stars seen within the instability strip have not been
 observed to pulsate despite having similar properties to known roAp stars.
 These stars, where pulsations are not observed, are commonly called the
 non-oscillating Ap (noAp) stars. Many attempts have been made to determine
 systematic differences between roAp and noAp stars (see the discussions in
 Hubrig et al. \cite{hubrig00} and Cunha \cite{cunha02}).  Hubrig et al. (\cite{hubrig00}) discovered that no roAp star is known to be a spectroscopic binary. 
Also, in the similar colour range to
 roAp stars, the noAp stars seem to be more evolved on average than the roAp
 stars (North et al. \cite{north97}), although this may be a selection
 effect (Elkin et al. 2005). This apparent distinction in the evolutionary
stages between roAp and noAp stars is supported theoretically by Cunha
 (\cite{cunha02}), who showed that a magnetic field is less likely to
 suppress convection in more evolved stars, and that the growth rate of
 unstable modes decreases as these stars evolve, so that evolved stars are
 stabilized more easily. This brief review shows how theory and observations
 are complementary to our understanding of the physics of these stars.

As far as the $\delta$\,Sct stars are concerned, these are in general
 chemically normal stars, either because the pulsations disrupt the element
 separation due to the diffusion process, or because diffusion stabilizes
 the star. It was long thought for these reasons that metallicism and
 pulsations were mutually exclusive. In the late 1970s, theoretical models
 predicted however that pulsations could occur in {\it{evolved}} Am stars:
 when these stars evolve, the He \textsc{ii} ionization is shifted more
 deeply into the star, where there is sufficient residual He to drive
 pulsations (Cox, King \& Hodson \cite{cox79}). Such cases of $\rho$\,Pup
 stars\footnote{The group of $\rho\,Puppis$ (formerly $\delta\,Delphini$)
 stars are luminous, cool, evolved stars that may exhibit low-overtone
 pulsation characteristic of $\delta$\,Sct variability (Kurtz
 \cite{kurtz76}; Turcotte et al. \cite{turcotteetal}).} are now not unusual.
 Extreme cases of co-existence of metallicism and long-period pulsations
 have been found in the evolved Am star HD\,40765 (Kurtz et al.
 \cite{kurtz95}) and in the strongly peculiar star HD\,188136 (Kurtz 1980;
 Wegner 1981). The present paper briefly discusses the existence of two other evolved,
 long period Am pulsators, HD\,98851 and HD\,102480, also discovered during this survey
and published Joshi et al. (\cite{joshi03}).

Cunha (\cite{cunha02}) predicted that longer period oscillations (20-25 min)
in magnetic roAp stars should exist in the more evolved stars. The works of
Turcotte (\cite{turcotte00}) and Turcotte et al. (\cite{turcotteetal}) also
support this theory. Periods longer than about 16\,min remained for a long
time not observed in the known roAp stars, which are concentrated in the
lower (fainter) part of the theoretical instability strip; in fact,
$\delta$\,Sct type of pulsations were thought to be completely suppressed in
roAp stars (Kurtz \& Martinez \cite{kurtzm00}). Recent works still support
this view, and show theoretically that both low-order adiabatic (Saio \&
Gautschy \cite{saio04}) and non-adiabatic (Saio \cite{saio05}) oscillations
are damped by the magnetic field. Recently, however, Elkin et al.
(\cite{elkin05}) discovered pulsations in an evolved and luminous Ap star
HD\,116114, with a pulsation period of $21$\,min -- the longest period among
the known roAp stars. This in turn may support the theory that longer
periods do exist in luminous stars, and that the current observational
distinction between the roAp and noAp stars are due to an observational bias
toward fainter stars. Also, HD\,21190 was reported by Koen et al.
(\cite{koen01}) to be both a very evolved Ap and a $\delta$\,Sct star. The
classification of this star is F2III\,SrEuSi:, which indicates a possible
magnetic nature. Since some works ({\it{e.g.}} Saio \cite{saio05}) showed
theoretically that the presence of a magnetic field stabilizes the star
against low-overtone $\delta$\,Sct pulsation, the case of HD\,21190 is
particularly interesting. Beyond this particular star, the search of
low-overtone $\delta$\,Sct pulsations in magnetic Ap star is very important
for our understanding of magnetism and pulsations.

We note here that for the detection of pulsation, the method of spectroscopic studies of radial velocity variations from time series can  be more efficient than the photometric studies,  as illustrated by the cases of HD\,116114 (Elkin et al. \cite{elkin05}),  HD\,154708 (Kurtz et al., in preparation) and $\beta$ CrB (Hatzes \& Mkrtichian \cite{hatzes04}).  See also the review of Kurtz (\cite{kurtz05}) on this topic.  

For main sequence stars, the coexistence of spectral peculiarity and
pulsations for a long time remained doubtful. Kurtz (\cite{kurtz78},
\cite{kurtz84}) first found low-amplitude (a few mmag) $\delta$\,Sct
pulsation in the marginal Am stars HR\,4594, HR\,8210 and HR\,3321. Kurtz
(1989) further reported the discovery of the classical Am star HD\,1097 as a
$\delta$\,Sct star, thereby showing that both classical and evolved Am stars
can pulsate. Two more classical Am pulsators were discovered in the
framework of the present survey: HD\,13038 and HD\,13079; see Paper I.

The more detailed the analysis of pulsations in CP stars, the more
complicated the picture. HD\,188136 is, for instance, a very peculiar,
multiperiodic $\delta$\,Sct star with $50$-mmag peak-to-peak variations
(Kurtz \cite{kurtz80}). The extreme peculiarity of this star is surprisingly
not mixed away by the large amplitude pulsations. It is classified as a
$\delta$\,Del ($\rho$\,Pup) star, but the Rare Earth abundances in this star suggest the possible presence of a strong magnetic field (Wegner \cite{wegner81}) --
which, if true, would mean that magnetic fields do not always prevent
$\delta$\,Sct type of pulsations.

In some stars the $\delta$\,Sct pulsations may exhibit both p~modes and g~modes (e.g. HD\,50018; Zhiping \cite{zhiping00}). Henry \& Fekel (\cite{henry05}) have shown that HD\,8801 appears to be an Am star that is also both a $\delta$\,Sct star and a g-mode $\gamma$\,Doradus star (see Henry \& Fekel \cite{henry03} for a review of $\gamma$\,Dor stars).

In many respects, the information that we have on pulsations and metallicity
is therefore a puzzle. What we observe (and what the theoretical models can
at least partially explain) is that evolved Am stars -- i.e. $\rho$\,Pup, or
$\delta$\,Del stars -- can pulsate (i.e., the $\kappa$-mechanism in
He\,\textsc{ii} can drive the pulsation in evolved stars, and low-amplitude
pulsations may not mix away the peculiarities); that marginal Am stars can
be low-amplitude $\delta$\,Sct stars (a sufficient amount of He remains to
drive low-amplitude oscillations) and that strong magnetic fields and high-overtone pulsation can, and actually do, coexist in the roAp stars.

Among the things still to be ascertained are the mechanisms making some classical Am stars $\delta$\,Sct pulsators (in particular why large amplitude pulsations do not mix away the peculiarities), how $\delta$\,Sct pulsations may occur in the presence of strong magnetic fields, and how frequent is the occurrence of long period pulsations in evolved roAp stars. As in the past, much insight into these questions will most probably come from observational discoveries. 

Since the early 1980s, the South African working group has devoted a lot of
 time to detect mmag variations in Ap and Am stars. They discovered more
 than twenty new roAp stars (Martinez \& Kurtz \cite{martinez95}) and
 published an extensive list of null results (Martinez \& Kurtz
 \cite{martinez94}). Of the 31 roAp stars known in 1997, only 3 were in the
 northern sky hemisphere. This situation gave birth to a number of surveys
 in the northern hemisphere to find new members of this group (Heller \&
 Kramer \cite{heller98}; Nelson \& Kreidl \cite{nelson93}; Dorokhova \&
 Dorokhov \cite{dorokhova98}; Handler et al. \cite{handler99}). The
 Nainital-Cape Survey is one of them, and as such one of its major
 motivations is to search new roAp stars in the northern hemisphere.
 Regarding the discussion above, this kind of survey presents numerous other
 potential benefits: basically, they represent the primary materials to
 determine statistically the physical properties of the complicated classes
 of stars described above, and to constrain more precisely the theoretical
 models. More precisely, the instrumental setting of the present survey for
 acquiring high-precision, fast photometric observations (see
 Sec.~\ref{obs}) is particularly well suited to search for long periods
 (above $\approx$ 20 min) in magnetic Ap stars (cf. Turcotte et al.
 \cite{turcotteetal}; Cunha \cite{cunha02}; Saio \cite{saio05}) to search
 for low-overtone $\delta$\,Sct pulsations in (either cool or evolved) A-F
 peculiar stars, and to search for the coexistence of p- and g-modes
 pulsations in Am stars. The selection of the candidates was traditionally
 oriented towards cool objects (Martinez \& Kurtz \cite{martinez94}) but we
 include in our list evolved stars as well (see Sec. \ref{targets}).

 The rest of the paper is organized as follows: Section 2 summarizes the strategy for the selection of the candidates, the observations and the data analysis procedure. Section 3 presents five variables discovered during this Survey. The null results are presented and analysed in Section 4. The last section summarizes our results. 

\section{Selection of the candidates, observations and data reduction}

\subsection{Selection of the candidates}\label{targets}

In order to increase the chances of discovering variability in Ap and Am stars, the strategy adopted for the current survey was to select candidates presenting Str\"{o}mgren photometric indices similar to those of the known variable Ap and Am stars (see Paper I). The range of the indices was also slightly extended with respect to Paper I to take in the evolved stars that might be peculiar. The primary source of candidates for the survey was from Str\"{o}mgren photometry of A and F-type stars in the Simbad data base (Hauck \& Mermilliod \cite{hauck98}). The following range of Str\"{o}mgren photometric indices was used to select candidates: $0.46 \le c_1 \le 0.88$; $0.19 \le m_1 \le 0.33$; $2.69 \le \beta \le 2.88$; $0.08 \leq b-y \le 0.31$; $-0.12 \le \delta m_1 \le 0.02$ and $\delta c_1
\le 0.04$, where $c_1$ is the Balmer discontinuity parameter, an indicator of luminosity; $m_1$ is the line-blanketing parameter, an indicator of metallicity; $\beta$ is the $H_{\beta}$ line strength index, reasonably free from reddening, an indicator of temperature in the range from around A3 to F2; $b-y$ is also an indicator of temperature, but is affected by reddening. A more negative value of $\delta m_1$ indicates a stronger metallicity; a negative $\delta c_1$ index is an indicator of peculiarity. For the more evolved, more luminous stars $\delta c_1$ may be positive, even for strongly peculiar stars, since $c_1$ increases with luminosity.

Another indicator for peculiarity is the ``$\Delta p$ parameter'' as defined by  Masana et al. (\cite{masana98}). According to these authors, 50\% of the late region CP stars (roughly A2 and later)  should  present a $\Delta p$ greater than  $\Delta p_0 =2$\,mag, and only 17\% of normal stars should present a $\Delta p$ greater than 2\,mag (see eq. (5) of their paper; for hotter stars eq. (4) should  be used, and interpreted with  $\Delta p_0 =1.25$\,mag). Most of our star have $\Delta p$ scattered around 2\,mag, and $\Delta p > 2$\,mag for 60 stars (see Table 1). The mean and the median of the distribution of the $\Delta p $ are respectively 2.08\,mag and 1.99\,mag,  indicating strong peculiarity for those stars. The combination of the spectral types, Str\"{o}mgren photometric and $\Delta p$ indices demonstrates chemical peculiarity in almost all of the stars we have studied. Some normal stars and some hotter stars have been included to extend the range of our search, and for comparison. 

\subsection{Observations and data reduction}\label{obs}

For the Nainital-Cape Survey, high-speed photometric observations of Ap and
Am star candidates are carried out from ARIES (Manora Peak, Nainital) using
a three-channel fast photometer attached to the ARIES 104-cm Sampurnanand
telescope (Sagar \cite{sagar99}; Ashoka et al. \cite{ashoka00}). Most of the
selected program stars are between 6 and 10 mag. It is therefore very
difficult in general to find a nearby comparison star of similar magnitude
and colour as those of the program star. Hence, most of the time we use the
photometer in two-channel mode (one channel measuring the target star plus
sky background and a second channel measuring sky only). The time-series
photometric observations consist of continuous $10$-s integrations obtained
through a Johnson $B$ filter. This filter is expected to yield the highest
amplitude variations and to maximize the number of counts. An aperture of
30$^{\prime\prime}$ is used to minimize flux variations caused by seeing
fluctuations and guiding errors. The observing protocol is simple. We acquire
time-series photometric observations of the candidate stars for $1$ to
$3$\,hr in order to be able to reveal both roAp and $\delta$\,Sct pulsations
on photometric nights. As a single null result is insufficient to exclude a
candidate from being variable, several runs are recorded for some of the
same target stars (see Sec. 4).

The data reduction process comprises the following steps: (a) Visual
inspection of the light curve to identify and remove all obvious bad data
points; (b) correction for coincident counting losses; (c) subtraction of
the interpolated sky background; and (d) correction for the mean atmospheric extinction
($\langle \kappa_B \rangle = 0.26$ for Nainital; the best extinction
coefficient is calculated for each observing run). After applying these
corrections, the times of the mid-points of the observations are converted
into heliocentric Julian dates (HJD) with an accuracy of $10^{-5}$ day
($\approx $1\,s). The reduced data comprise a time-series of the HJD and B
magnitudes with respect to the mean of the run. These data are then analyzed
using an algorithm based on the Discrete Fourier Transform (DFT) for
unequally spaced data (Deeming \cite{deeming75}; Kurtz \cite{kurtz85}). The
DFT of the time series produces an amplitude spectrum of the light curve.
Since we seldom can observe a comparison star, there is always some degree
of low-frequency sky transparency variations (see Sec. 4) mixed with the
possible low-frequency stellar variations. The sky transparency variations
are well-separated in frequency space from possible roAp pulsation
frequencies.

\section{Pulsating Variables Discovered During this Survey} 
The pulsating variables discovered since the last survey
report in Paper I are discussed below. The results about the first two stars below have been
already published: we  only summarize here the main results and discuss them.

\subsection{HD\,98851} 
HD\,98851 ($\alpha_{2000} = 11\,22\,51.17$; $\delta_{2000} = +31\,49\,41.1$; $m_B=7.72$; $m_V= 7.41$) is a star of spectral type F2. The Str\"{o}mgren photometric indices for this star are: $b-y=0.199$, $m_1=0.222$, $c_1=0.766$ (Hauck \& Mermilliod \cite{hauck98}).
There is no published value of the $\beta$ index. Using low resolution spectroscopy, Joshi et al. (\cite{joshi03}) found that this is a cool star of effective temperature 7000 K. The calibration of Crawford (\cite{crawford75}) for F-type stars gives $\delta m_1 = -0.051$ and $\delta c_1 = 0.236$. The $\delta m_1$ index is well within the range of Str\"{o}mgren photometric indices for the known roAp stars; $\delta c_1$ indicates that the star is evolved. Abt (\cite{abt84}) classified this star as a marginal Am star, with Ca \textsc{ii} K line, H\,\textsc{i} Balmer lines and metallic lines (respectively, K/H/M) of the types F1/F1 IV/F3.

Joshi et al. (\cite{joshi00}) found $\delta$\,Scuti pulsation with  three main frequencies, $f_1= 0.20$\,mHz, $f_2=0.10$\,mHz and $f_3=0.02$\,mHz. Alternating high and low amplitude cycles visible in the light curves indicate a nearly sub-harmonic period ratio of $2$:$1$, a phenomenon not commonly observed in other Am or $\delta$\,Sct stars (Joshi et al. \cite{joshi03}). Zhou (\cite{zhou01}) also observed this star photometrically from the Xinglong Station of the National Astronomical Observatory, China, using a three-channel fast photometer attached to a 85-cm Cassegrain telescope, and reported similar types of light curves with the same periodicity. 

\subsection{HD\,102480} 
The star HD\,102480 ($\alpha_{2000} = 11\,47\,52.88$; $\delta_{2000} =
+53\,00\,54.5$; $m_B=8.78$, $m_V= 8.45$) is of spectral type F1. Its
Str\"{o}mgren photometric indices are $b-y=0.211$, $m_1=0.204$, $c_1=0.732$
(Hauck \& Mermilliod \cite{hauck98}). The Str\"{o}mgren metallicity and luminosity indices are
$\delta m_1 = -0.034$ and $\delta c_1 = 0.292$. The effective temperature is
6750\,K derived using low-resolution spectroscopy (Joshi et al.
\cite{joshi03}). The K/H/M lines are also those of a marginal Am star:
Am(F2/F4/F4) (Abt \cite{abt84}). The Str\"{o}mgren photometric indices in
combination with effective temperature indicate that this is a cool, evolved
marginal Am star.
 
The photometric variability of this star was discovered from ARIES by Joshi
et al. (\cite{joshi02}). 
Combining data sets closely spaced in time, the amplitude spectra show that
HD\,102480 pulsates with three frequencies: $f_1=0.107$\,mHz, $f_2=
0.156$\,mHz and $f_3=0.198$\,mHz (Joshi et al. \cite{joshi03}).  As with HD~98851,
the light curves present alternating high and low amplitude
 variations with a period ratio close to 2:1.

 The light curves of HD\,98851 and HD\,102480 are similar to those of the
 luminous yellow supergiant pulsating variables RV Tauri stars, which show
 alternating deep (primary) and shallow (secondary) minima, with a periods
 in the range $30 – 150$\,d and a brightness range of up to four magnitudes.
 This group of stars is composed of binaries with circumstellar or
 circumbinary disks which interchange material with the photosphere. It is
 thought that the underlying variability in RV Tauri stars arises from
 pulsations, with the alternating light curve arising from a 2:1 resonance
 between the fundamental and first overtone modes.
The unusual nearly harmonic period ratios,
 alternating high and low maxima, and Am spectral types of HD~98851 and
HD~102480 make these stars particularly interesting objects for further observational and
 theoretical studies.

\subsection{HD\,113878} 
HD\,113878 ($\alpha_{2000}=13\,06\,00.74$; $\delta_{2000} = +48\,01\,41.3$;
$m_B = 8.59$, $m_V= 8.24$) is an early F-type star, with the following
Str\"{o}mgren photometric indices: $b-y=0.219$, $m_1=0.257$, $c_1=0.729$,
$\beta =2.745$ (Hauck \& Mermilliod \cite{hauck98}), giving $\delta m_1 =
-0.037$ and $\delta c_1 = 0.069$. From Abt (\cite{abt84}), the spectral type
of this star is F1/F3V/F3 indicating that HD\,113878 is a marginal Am star.

\begin{figure}[h] 
\includegraphics[width=.5\textwidth]{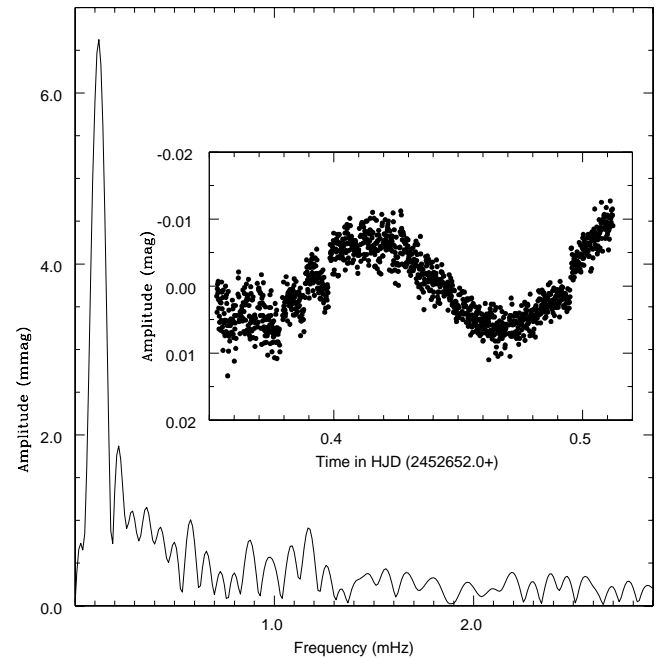} 
\caption{Johnson $B$ light curve (inset) and corresponding amplitude spectrum of HD\,113878 obtained on HJD2452652.} 
\end{figure} 

We have discovered low-amplitude pulsations in HD\,113878 (Fig.\,1). From
the analysis of the light curves obtained on different nights, the star has
a clear variability period of $2.31$\,hr ($f_1=0.12$\,mHz), which is typical
of $\delta$\,Sct stars. To investigate possible additional pulsational
frequencies, we observed this star for a total of 8 photometric nights.
Unfortunately, the large time gaps between the runs made the identification
of the additional frequencies particularly difficult because of frequency
aliases in the Fourier spectra. In order to study this star in more detail, densely sampled
time-series data are required.
 
Very interestingly, Abt (\cite{abt84}) showed that the stars HD\,98851,
HD\,102480, HD\,113878 and HD\,104202 are similar in the sense that they do
not show the classical difference of more than $0.5$ spectral class between
the K line types and the metallic lines. These four stars are, however,
classified as Am stars by their strong Sr\,\textsc{ii} lines, weak
$\lambda$4226 Ca\,\textsc{i} line, and by  Str\"{o}mgren $\delta m_1$ indices
supporting  their peculiarity.
 Among these stars, HD\,98851, HD\,102480 and HD\,113878 turned
out to be photometric variables. 
HD\,98851 and HD\,102480 also exhibit interesting light curves with
alternating high- and low-amplitude maxima. If
the star HD\,104202 turns out to exhibit variability as well, then
these objects may provide a useful insight on the relationship of chemical
peculiarity to variability.

\subsection{HD\,118660} 
HD\,118660 ($\alpha_{2000} = 13\,38\,07.89$; $\delta_{2000} =
+14\,18\,06.9$; $m_B = 6.746$: $m_V = 6.500$) is a bright star of spectral
type A8V (Abt \& Morrell \cite{abt95}). Its Str\"{o}mgren photometric
indices are $b-y = 0.150$, $m_1 = 0.214$, $c_1 = 0.794$ and $\beta = 2.778$
(Hauck \& Mermilliod \cite{hauck98}), giving $\delta m_1 = -0.018$ and
$\delta c_1 = 0.054$, indicating that the star is metallic and evolved.

\begin{figure}[h] 
\includegraphics[width=.5\textwidth]{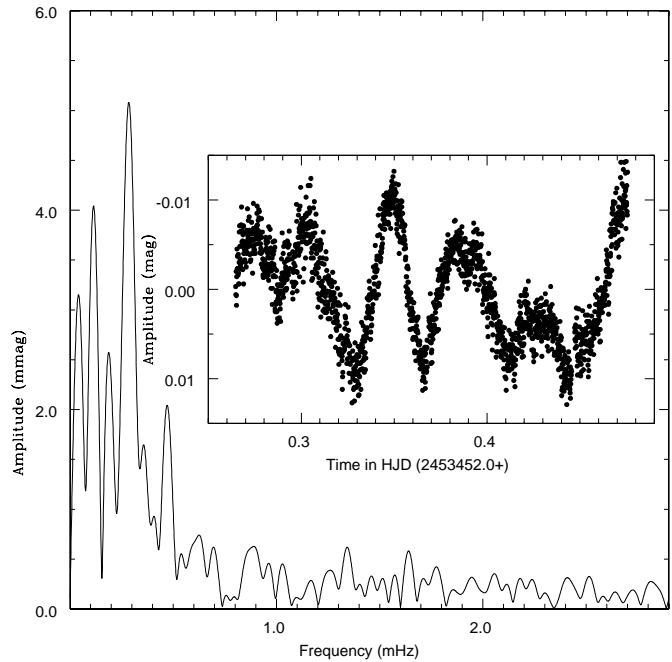} 
\caption{Johnson $B$ light curve (inset) and amplitude spectrum of HD\,118660 obtained on HJD2453452.} 
\end{figure} 

Time-series data (Fig.\,2) have revealed that this star is a multi-periodic
$\delta$\,Scuti variable pulsating with a principal period of about 1\,hr.
Another prominent period of 2.52\,hr is also apparent. Preliminary analyses
indicate that many other periods may also be present in the data, though we
could not investigate them further so far. More time-series data are
consequently required. At a declination of about 14$^\circ$, HD\,118660 is a
good target for a multi-site campaign with 0.5-m to 1.0-m class telescopes
situated in both northern and southern hemispheres.

\subsection{HD\,207561} 
HD\,207561 ($\alpha_{2000} = 21\,48\,16.05$; $\delta_{2000} = +54\,23\,14.6$; $m_B = 8.09$; $m_V = 7.85$) is an F0III star with Str\"{o}mgren indices of $b-y = 0.142$, $m_1 = 0.220$, $c_1 = 0.820$ and $\beta = 2.825$, giving metallicity and luminosity indices of $\delta m_1 = -0.014$ and $\delta c_1 = -0.040$. 
These indices are typical for roAp stars, but this star is classified as an Am star (Cowley \& Cowley \cite{cowley65}; Bertaud \& Floquet \cite{bertaud74}; Nicolet \cite{nicolet82}). 

We have discovered 6-min oscillations in the light curve of this star on two
consecutive nights, as shown in Fig.\,3 and 4. For this first two nights,
the variability is evident.
Fig.\,5 shows the amplitude spectra for the light curves on these three
nights where a peak at 2.75\,mHz ($P = 6.1$\,min) is clear with excellent
signal-to-noise ratio on both nights. Both the light curves and amplitude
spectra are typical of the roAp stars. In our experience studying these
stars for more than 20 years, we have never seen a spurious signal as clear
as this for any star. Furthermore, in the next section we show the null
results from our survey where no signal stands out like this one. This shows
that our instrumentation and sky conditions do not produce false peaks such
a high frequency as we see in Fig.\,5.

Thus it seems that HD\,207561 is a newly discovered roAp star. However, we
have been unable to reproduce the results shown in Figs\,3 and 5 over two
following seasons. We have observed this star for a total of 43.5\,hr on 17
nights over three observing seasons. On some nights there is no signal at
all to high precision, as we show in Fig.\,4 and in the bottom panel of Fig.\,5. On
other nights we have seen possibly significant peaks, but at lower
frequencies than the one detected in Figs\,3 and 5.

While roAp stars are oblique pulsators, so their observed amplitude varies
with rotation and can decrease to zero, or rise to many mmag from
night-to-night, most of them have stable frequency spectra. Exceptions to
this is HD\,60435 (Matthews et al. \cite{matthews87}) where the mode
lifetimes appear to be shorter than one week, and HD\,217522 which had a new
frequency at 2.02\,mHz in 1989 in addition to the 1.2\,mHz peak known from
1982 (Kreidl et al. 1991). We have not previously seen the apparent
behaviour that the light curves and amplitude spectra for the 17 nights of
observations of HD\,207561 show, so we do not claim with certainty that it
is an roAp star, although we strongly suspect this from the evidence shown.

From the Str\"{o}mgren indices given above, HD\,207561 is clearly a
chemically peculiar star -- either Am or Ap. These can be confused with each
other at classification dispersions, so the Am classification of the star
needs to be re-examined 
in light of the possible pulsation. A definitive
spectral type is needed, as is a search for the presence of a magnetic
field.
\begin{figure}[htpb]
\vspace{-1cm}
\centering 
\includegraphics[width=.5\textwidth]{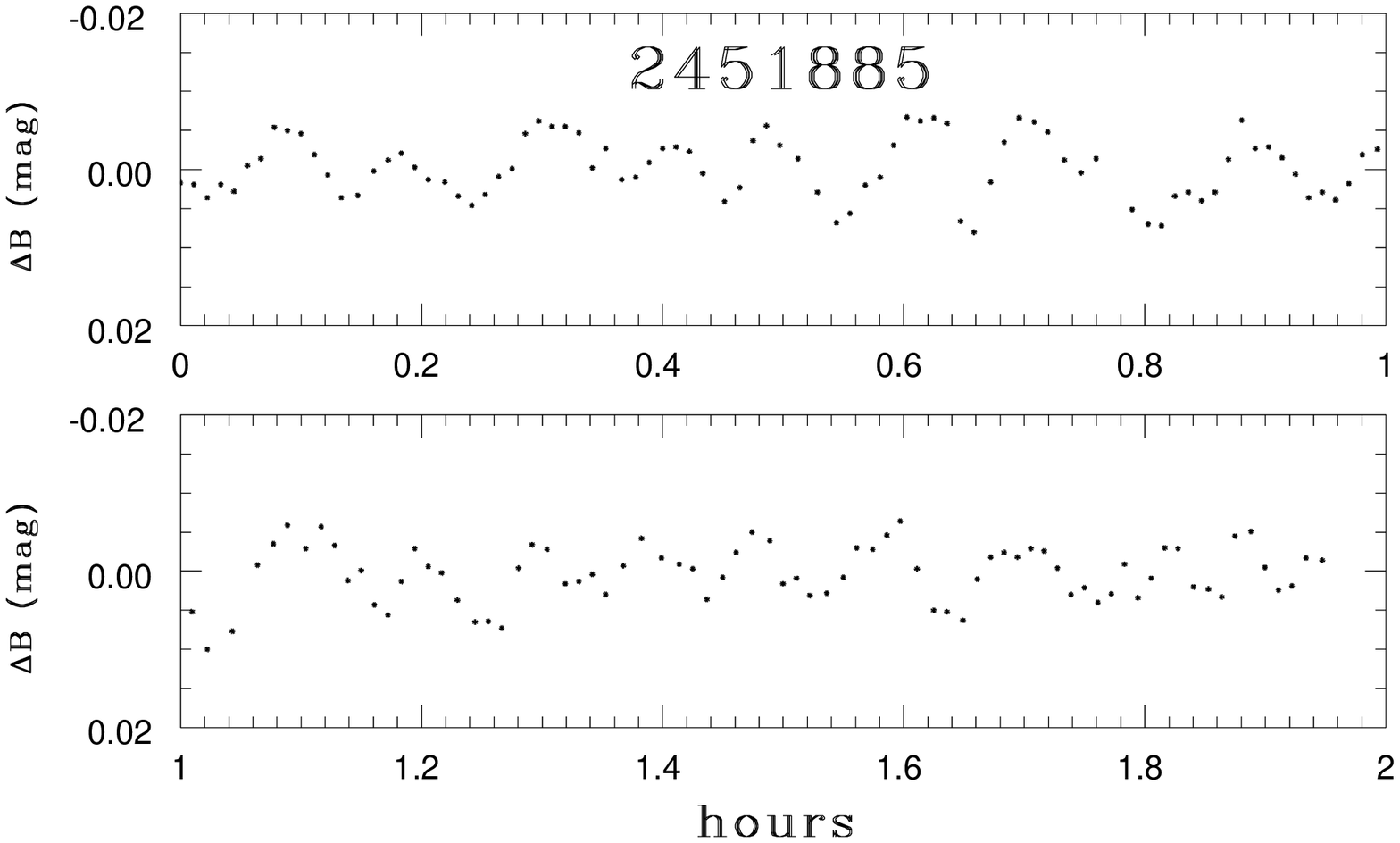}
\vspace{-3cm}
\vspace{-1cm}
\centering 
\includegraphics[width=.5\textwidth]{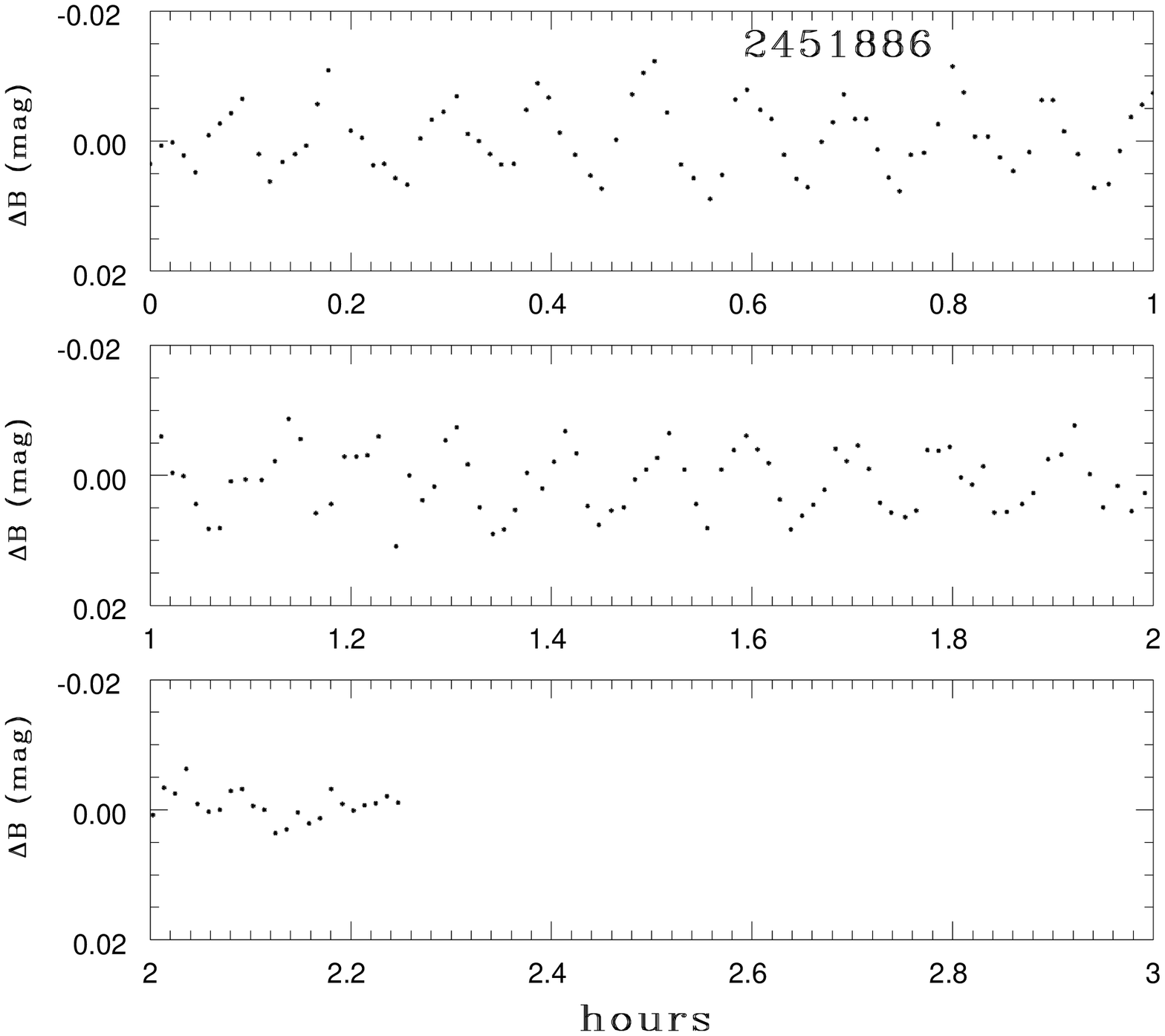}
\caption{Johnson $B$ light curves of HD\,207561 obtained on two consecutive nights.
 These curves  clearly show the 6-min oscillations.  } 
\end{figure}
\begin{figure}[htpb]
\vspace{-3cm}
\centering 
\includegraphics[width=.5\textwidth]{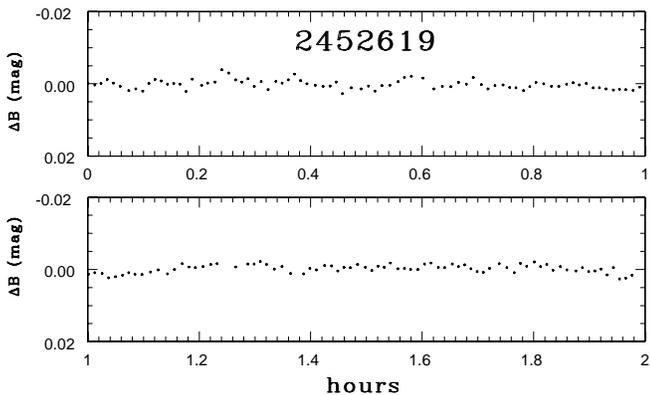}
\caption{Johnson $B$ light curves of HD\,207561 obtained on a
 third night from a different season. This curve 
shows a clear lack of signal at high precision. In all light the curves, the data have been merged to 
40-s integrations and low-frequency sky transparency variations have been 
prewhitened at frequencies below 0.5\,mHz.}
\end{figure} 

\begin{figure}[htpb]
\centering 
\includegraphics[width=.5\textwidth]{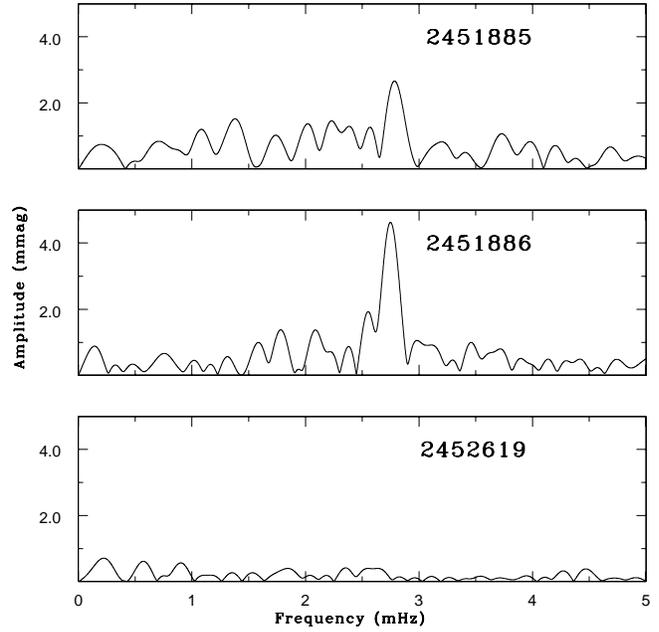} 
\caption{Amplitude spectra of the light curves shown in Fig.\,3 and 4 (HJDs are mentioned in each panel). The peaks at 2.75
\,mHz in the top two panels have high signal-to-noise ratio. We have never seen a case where signals as strong as this, and at this high frequency, have been 
instrumental or atmospheric in origin. We therefore consider them to be real, but have not been able to confirm them on 15 other nights of observation. Note, 
however, that roAp stars can be amplitude variable due to rotation and 
multiperiodicity. See the text for further discussion.} 
\end{figure}

If the 6-min pulsations in HD\,207561 can be confirmed, this star will be
interesting to study for its amplitude variability to determine if it is due
to short mode lifetimes or rotation. We expect that it will be found to be
an Ap star; if it should be an Am star, then the short period is unique and
demands extensive further study. Unfortunately, this star is poorly placed
for observing from ARIES. It is only observable during October to December
from this site because the monsoon precludes observing it earlier than
October. The best signal-to-noise ratio for the observations is obtained
when the sky is stable and the object is near the meridian. The sky
transparency usually becomes stable only two to three hours after the sunset
at Nainital. But by that time, HD\,207561 is two to three hours west of the
meridian. This makes it difficult to obtain long observing runs of this star
from Nainital. Consequently, observations of HD\,207561 from other astronomical sites would
be very useful to confirm the above results.
\section{Null Results}

As we shall see, the detection of small amplitude variations (a few mmag, or
less) in roAp stars, as well as low-amplitude $\delta$\,Sct pulsators, is
challenging. Detection of mmag pulsations can, however, be accomplished with
good conditions from a single site (down to about 0.3\,mmag for the highest
noise peaks, or a precision of about 0.1\,mmag in 1\,h of observing; see
Martinez \& Kurtz \cite{martinez94}), provided that scintillation noise is
sufficiently low, and that variations in the sky transparency are
sufficiently slow\footnote{For comparison, Kurtz et al. (\cite{kurtzetal05})
achieved 14-$\mu$mag photometric precision for the roAp star HR\,1217 with a
three-week multi-site campaign using the Whole Earth Telescope.}. Since the
stars of our survey are brighter than $m_V = 10$, sky transparency
variations and scintillation noise are the limiting factors in detecting the
small amplitude pulsations we are searching for.

We present here null results from our survey for 140 stars. The observations
cover a time range from 1999 January to 2004 December. They are presented in
both tabular and graphic form, and analyzed below. Table~1,
Fig.\,\ref{fig10} and Fig.\,\ref{fig11} show examples of the results for a
few stars; the complete table and figures are available in electronic form
with this paper. 

The twelve columns of Table~1 list respectively, for each
star: HD number, right ascension $\alpha_{2000}$, declination
$\delta_{2000}$, visual magnitude $V_m$, spectral type, spectral indices
$b-y$, $\delta m_1$, $\delta c_1$, $\beta$,  peculiar parameter $\Delta p$ 
(Masana et al. \cite{masana98}),  Julian dates on which the
star was observed, and time of each observing run in hours. Fig.\,\ref{fig10}
shows examples of spectra corrected for extinction, but filtered for
low-frequency noise that we consider to be mostly, or completely caused by
sky transparency variations. Fig.\,\ref{fig11} presents prewhitened spectra
that have been filtered for low-frequency noise. The prewhitening strategy
was to remove the low-frequency peaks in the range $0 -
0.5$\,mHz, if those were higher than peaks above $0.5$ mHz. Because of this
procedure, the amplitude threshold above which low frequency peaks were
removed varies from star to star. In this sense, the spectra corrected only
for extinction form a more homogeneous set of data, since the data reduction
process is the same for all stars. On the other hand, the corresponding
spectra often present a large low frequency peak, in the sidelobes of which
stellar peaks may be buried. We decided to present the two sets of data
because they are complementary, and both are required for a proper analysis
of the observations.

By observing the amplitude spectra of Fig.\,\ref{fig10}, one can notice that
the average level of the peaks is in general higher in the low frequency
region, and decreases towards higher frequencies. This higher level is
mainly caused by residual sky transparency variations. Above the
low-frequency region, the spectrum flattens to a level which corresponds to
the scintillation noise. This tells us that the detection of a peak in the
low-frequency region is more difficult than at higher frequencies because
the noise level is higher. Hence, detecting $\delta$\,Sct pulsations (longer
than about 0.5\,h) is difficult, unless they have very large amplitude. This
is the reason why long pulsation periods are difficult to detect with our
high-speed technique, and might explain why such pulsations are actually not
detected in evolved roAp stars. In the prewhitened spectra
(Fig.\,\ref{fig11}), note that the tallest peaks are not found in the
$[0-0.5]$\,mHz frequency region, because prewhitening has removed the
largest low frequency contributions. Prewhitening is aimed at removing the
major part of the sky transparency variations, but this process also removes
any pulsations which may be hidden at low frequency.

In this regard, the null results mean that no peak in any region of the
spectrum was sufficiently high with respect to the surrounding noise level
to be considered as the signature of a pulsating star. One may be tempted to
extrapolate that for any particular star which belongs to these null
results, there is at any given frequency no pulsation higher than the noise
level reported in the corresponding spectrum. This is true for the
particular epochs at which the star was observed, but it may {\it{not}} be
the case at other epochs. Rotational amplitude modulation and beating
between frequencies may result in pulsations not being visible at a
particular time (see, e.g., Martinez \& Kurtz \cite{martinez94} and Handler
\cite{handler04} for further discussions of these phenomena). It is
therefore likely that some of these 140 stars are actually pulsating. Many
of the stars have been observed only once or a very few times (see full
electronic figure and the Table). Hence, many of the corresponding results are
inconclusive rather than negative, and observers should not be discouraged
from observing these stars further.

\begin{figure}[t] 
\centering 
\includegraphics[width=.5\textwidth]{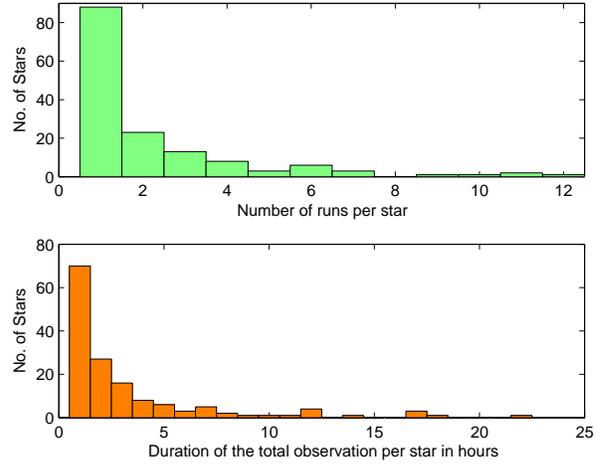} 
\caption{Statistics of the runs for the Nainital-Cape Survey. Top: Distribution of the number of runs per star. Clearly, most of the stars have been observed only once. A few of them, where some variability was suspected, were observed many times. Bottom: Distribution of the total time dedicated to observe one star (in hours). Most of the stars have been observed less than 2\,h. The median value is 1.6\,h and the mean 3.2\,h.} 
\label{F4}
\end{figure} 

For stars with multiple observing runs, if a null result is found every time the star is observed, no variability or variability with an amplitude below the noise level are strongly favoured conclusions. For those stars where variability could not be detected several times, this result is interesting, showing that there are stars lying inside the classical instability strip that do not show any variability at high precision. This knowledge is essential to our understanding of what distinguishes the pulsating CP stars from similar non-pulsating CP stars, and helps us to identify or refine the parameters which distinguish constant from variable CP stars.

For those stars that may in the future be found to be pulsating, even though classified as null results here, the present spectra provide a useful reference about the time and the sensitivity for which pulsations were not apparent. This was, for example, the case for HD\,116114, which was observed in two photometric searches for new roAp stars (Nelson \& Kreidl \cite{nelson93}; Martinez \& Kurtz \cite{martinez94}). No photometric variability was found in either survey, and the star was considered to be a noAp star. However, Elkin et al. (\cite{elkin05}) later discovered in HD\,116114 to have radial velocity variations with a period near 21\,min.

\begin{figure}[t] 
\centering 
\includegraphics[width=.5\textwidth]{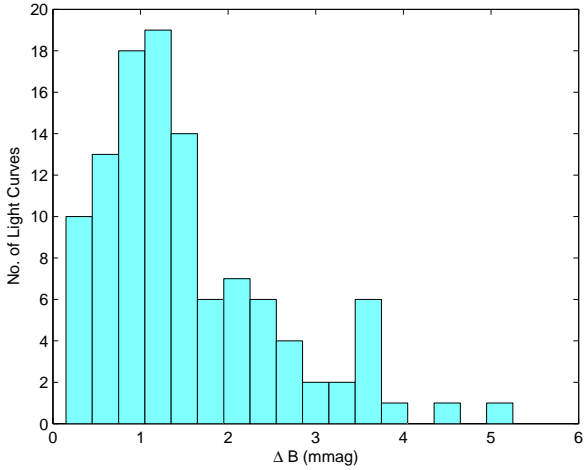} 
\caption{Distribution of the detection limits (as the tallest peak in the spectrum) in the Nainital-Cape Survey.} 
\label{F1}
\end{figure}

We shall go a bit further in the analysis of our null results, and turn to their detection limits. Assessing a criterion to derive a detection limit, with the purpose that this limit is both general and significant is difficult (see Mary \cite{mary05} and references therein for the problem of asteroseismic light curve analysis,  and Mary \cite{mary06} for the particular problem of the detection limits in fast photometry). The reasons are that the level of the sky transparency variation, that of the scintillation noise, and the frequency regions where either of these noise sources dominates varies from night to night, and sometimes during a given night, as well. For each particular run, one can assess the detection limit by considering the shape of the spectrum, the noise level in the flat frequency region, and the tallest peak of the spectrum. This tallest peak can be considered as an upper value of the detection limit for a particular observation. 
For each prewhitened light curve of our null results, we selected the tallest peak; the distribution of these detection limits is displayed in Fig.\,\ref{F1}.

\begin{figure}[t] 
\centering 
\includegraphics[width=.5\textwidth]{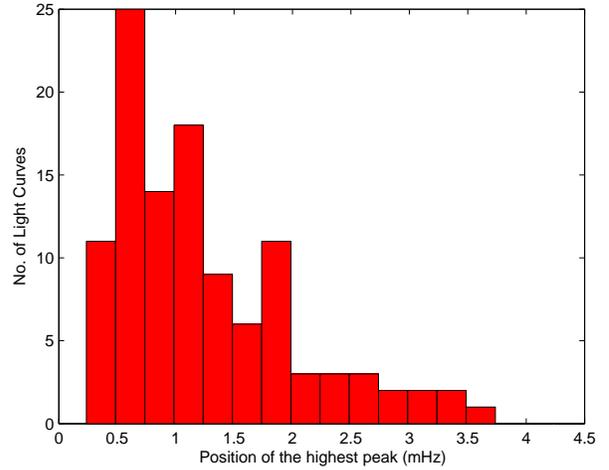} 
\caption{Distribution of the positions of the tallest peaks in the prewhitened light curves.} 

\label{F2}
\end{figure}

\begin{figure}[h]
\centering
\includegraphics[width=\textwidth]{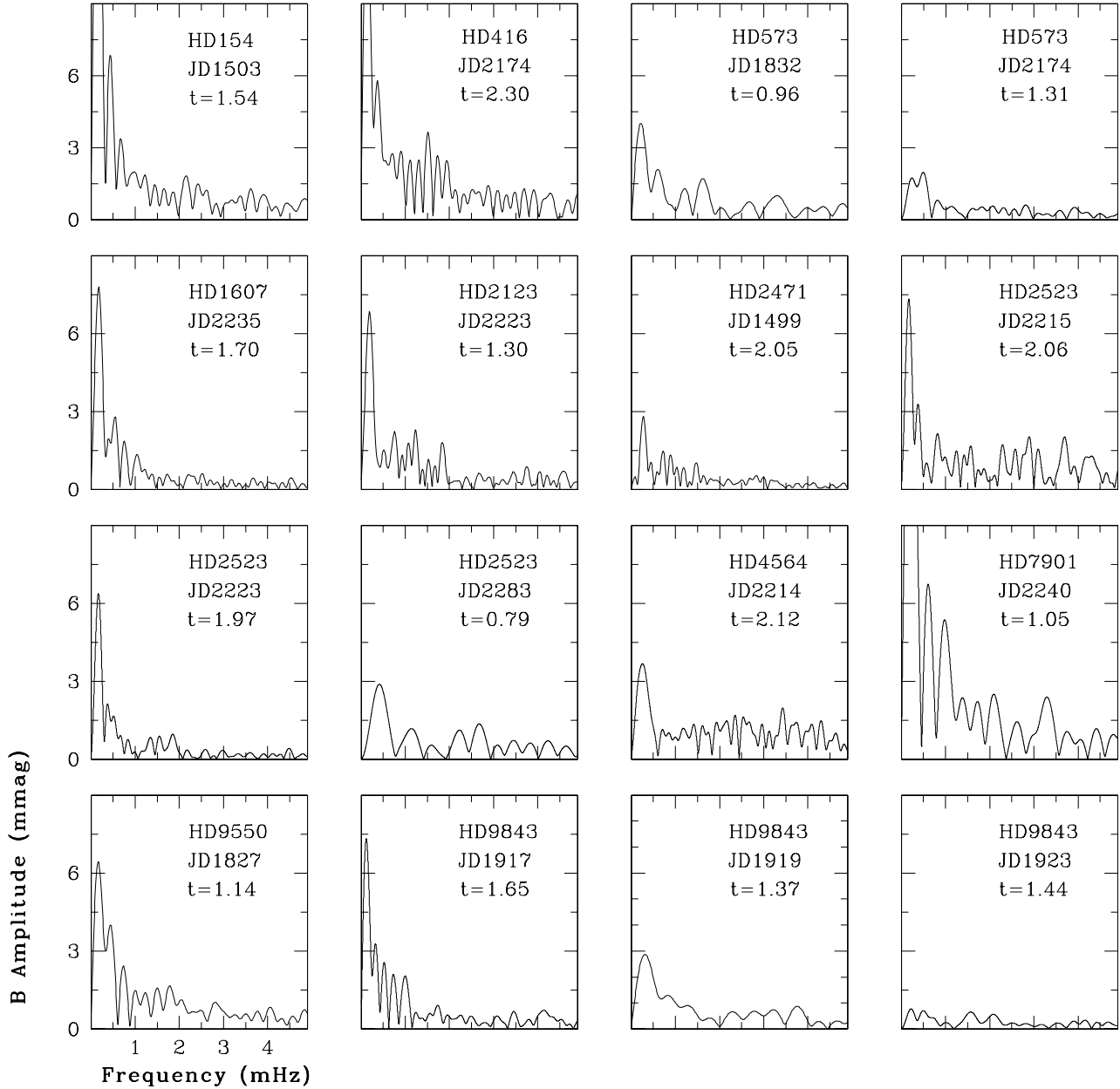}
\onecolumn
\caption{Null results from the Nainital-Cape Survey: Examples of amplitude spectra for
eleven sample stars 
corrected only for extinction, and in some cases for some very long-term sky
transparency variations. Each panel contains the Fourier transform of an individual light curve, covering a frequency range of $0$ to $5$\,mHz, and an amplitude
range of 0 to 9 mmag. The name of the object, date of the observation in Julian date (JD 245000+)
and duration of the observations in hours (hr), are mentioned in each panel. The rest of 
these amplitude spectra are available electronically. }
\label{fig10}
\end{figure}

\begin{figure}[h]
\centering
\includegraphics[width=\textwidth]{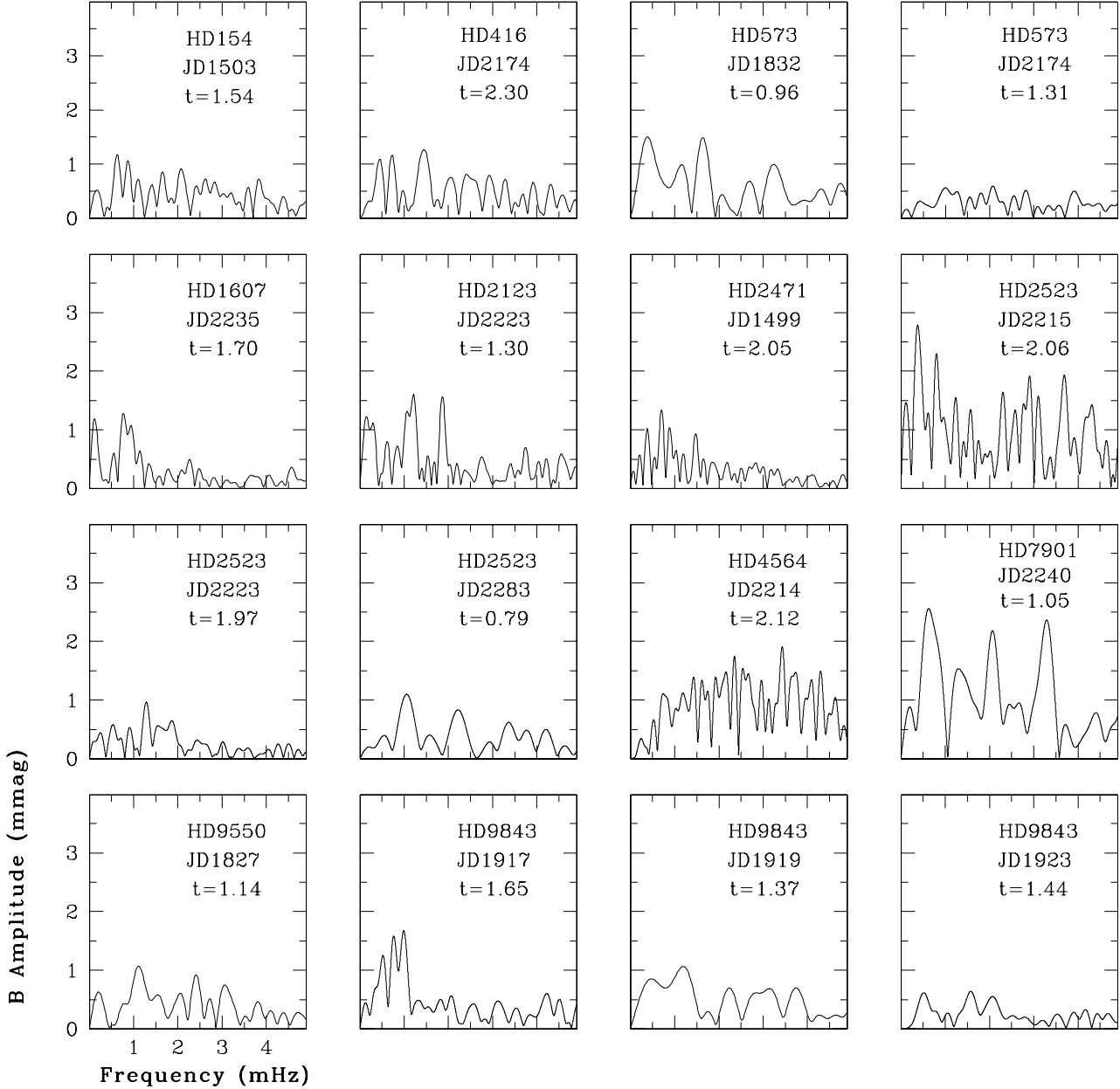}
\caption{Null results from the Nainital-Cape Survey: Examples of prewhitened
amplitude spectra for
eleven 
sample stars. Each 
panel contains the Fourier transform of an individual light curve, covering a frequency 
range of $0$ to $5$\,mHz, and an amplitude range of 0 to 4 mmag. The name of the object, 
date of the observation in Julian date (JD 245000+) and the length duration in hours (hr), are mentioned in 
each panel. The rest of the amplitude spectra are available electronically. }
\label{fig11}
\end{figure}

In most cases, the tallest peak has an amplitude in the range $0.5 -
1.5$\,mmag. For this distribution, the mean and median values are
respectively 1.5 and 1.3\,mmag; 1.5\,mmag, which is less than the largest
amplitude of many known roAp stars. On the other hand, the number of roAp
stars having small amplitude pulsations is expected to increase dramatically
below 1.5\,mmag, so that efforts towards pushing down this limit would most
probably lead to the discovery of many new variables. This may be
accomplished by increasing the duration of the observing runs for each star.
If we now turn to the frequency at which the tallest peak occurs, we obtain
the distribution of Fig.\,\ref{F2}.

Again, it can be observed that the prewhitening procedure has removed the
peaks at very low frequency.  Usually, the highest peak is found in the low
frequency region of $0.5 - 2$\,mHz; the median value is $1.0$\,mHz and the
mean $1.3$\,mHz. The occurrence of the tallest peak at these frequencies is
due to residual sky transparency variations, but probably also to undetected
$\delta$\,Sct and roAp stars. Above about 2\,mHz, the distribution tends to
flatten. It is not clear whether the bump around 1.9\,mHz in the typical
range for roAp star frequencies is caused by noise, or is indicative of more
pulsators in the sample. It could also be indicative of a low-level
instrumentally-induced frequency, such as a telescope drive oscillation,
affecting the data of many stars.

The above results are very similar to those obtained at the SAAO site and discussed by Martinez \& Kurtz \cite{martinez94} (see in particular their Figures~1 and 2 for a comparison of Fig.\,\ref{F1} and Fig.\,\ref{F2}). These analyses show in particular that Manora Peak (Nainital) is a good photometric site.

We shall make a few comments regarding the possible improvements of our null
results. As far as scintillation noise is concerned on the one hand, its
level can be reduced by increasing the size of the telescope and by

observing at a lower airmass (Young \cite{young67}; Dravins et al.
\cite{dravins98}). These two factors will be improved in the future since
ARIES is building a 1.3-m telescope 50\,km further away in the mountains,
and at a higher altitude (Devasthal site, 2420\,m instead of 1951\,m at
Manora Peak). There is also a project to build a 3-m telescope. For the
studies of this kind of variables, such a telescope will be useful for
photometry indeed, but also mostly for spectroscopy. The benefits of
spectroscopic studies for asteroseismology are huge; see, for instance, the
review of Kurtz (\cite{kurtz05}) on this topic, and the cases of HD 116114 (Elkin 2005, \cite{elkin05}) ,  HD154708 (Kurtz et al., in preparation) and $\beta$ CrB (Hatzes \& Mkrtichian \cite{hatzes04}).  As for reducing sky
transparency variations on the other hand, it would be possible to improve
the detection in the low frequency range by increasing the duration of
the runs, but this would be at the price of investigating a reduced number
of candidates -- a compromise to be dealt with in this kind of survey.

\section{Conclusions} 

We presented in this paper results obtained from the Nainital-Cape Survey
from the years 2000 to 2004.  We reported the pulsation discovered in several
stars with CP spectral classifications during this time range; we also presented
and analyzed our null results.

Pulsations of the $\delta$\,Sct type were discovered in the evolved Am stars
HD\,98851 (main frequencies: 0.20\,mHz and 0.10\,mHz) and HD\,102480 (main
frequencies: 0.09\,mHz and 0.19\,mHz). Very interestingly, both stars
exhibit unusual and alternating high and low amplitude maxima, with a period
ratio of 2:1. The star HD\,113878 pulsates with a period of $2.31$\,hr,
which is also typical for $\delta$\,Sct stars. The three stars HD\,98851,
HD\,102480 and HD\,113878 form part of a group of four stars described by
Abt (1984) as similar in the sense that they do not show the classical
difference of more than $0.5$ spectral class between the K line types and
the metallic lines. These four stars are, however, classified as Am stars by
their strong Sr\,\textsc{ii} lines, weak
$\lambda$4226 Ca\,\textsc{i} line, and other indications of their general
abnormality. The other star in this group is (HD\,104202). If that star
turns out to be variable as well, these objects can provide interesting
insights on the relationship of chemical composition and pulsations. The
star HD\,118660 seems to exhibit multi-periodic variability with a prominent
period of nearly $1$\,hr. Further observations are required to determine the
additional periods. This bright star is a good target for a multi-site
campaign using small telescopes. Finally, in HD\,207561, classified as an Am
star, an intriguing pulsation period of 6\,min was clearly detected in the
light curves obtained on two consecutive nights. It does not seem that this
oscillation is caused by instrumental noise, but this result should be taken
with care. The importance of the question of the coexistence of pulsations,
chemical peculiarity and magnetic field calls for further photometric,
spectroscopic, and magnetic studies of this star.

The Nainital-Cape survey is an on-going project. As the survey progresses,
new results are obtained regularly. In order to improve our strategy for the
current and future observations, more luminous peculiar stars were recently
included in our list. The peculiarity criterion defined by Masana et al.
(\cite{masana98}) is a useful tool in this direction. Up to now, the
results obtained from the survey (reported in Paper I and here) have brought
new and important information regarding the general properties of variable
Ap and Am stars. These results will contribute to, and stimulate other
investigations of these fascinating objects.

\begin{acknowledgements} 
SJ acknowledges the help from {\it INDU} for reading the manuscript rigorously. 
This work was carried out under the Indo-South African Science and
Technology Cooperation Program as a joint project titled 'Nainital-Cape Survey
for roAp stars,' funded by Departments of Science and Technology of the Indian
and South African governments.
\end{acknowledgements}

\begin{table*}[ht]
\begin{center}
\onecolumn
\caption[]{Sample of stars classified as null results during the Nainital-Cape survey. The unprewhitened and prewhitened spectra of these stars are depicted in Fig.\,\ref{fig10} and Fig.\,\ref{fig11}, respectively. The columns list: HD number, right ascension $\alpha_{2000}$, declination $\delta_{2000}$, visual magnitude $V$, spectral type, spectral indices $b-y$, $m_1$, $c_1$, $\beta$, $\delta m_1$, $\delta c_1$,  peculiarity parameter $\Delta p$ (Masana et al. \cite{masana98}), Julian dates (2450000+) on which the star was observed, and the duration of observation ($\Delta t$) in hours. The full table is available in the electronic material attached to this paper.
}
\vspace{1.0cm}
\begin{tabular}{rlllcrrrrrrrcc}
\hline
\multicolumn{1}{c}{ HD } & \multicolumn{1}{c}{$\alpha_{2000}$} & \multicolumn{1}{c}{$\delta_{2000}$} & \multicolumn{1}{c}{$V$} & \multicolumn{1}{c}{Spectral}& \multicolumn{1}{c}{$b-y$} & \multicolumn{1}{c}{$ m_1$}  & \multicolumn{1}{c}{$ c_1$} & \multicolumn{1}{c}{$\beta$} & \multicolumn{1}{c}{$\delta m_1$} & \multicolumn{1}{c}{$\delta c_1$}  & \multicolumn{1}{c}{$\Delta p$}  & \multicolumn{1}{c}{JD} & \multicolumn{1}{c}{ $\Delta t$}\\

\multicolumn{1}{c}{} & & & & \multicolumn{1}{c}{Type} & & & & & & & & \multicolumn{1}{c}{} & \multicolumn{1}{c}{hr}\\
\hline

 & & & & & & & & & & & & \\
154 & 00 06 24 & 34 36 25 & 8.93 & F0   & 0.223 & 0.230 & 0.746 & 2.782  & -0.033 & 0.002 & 2.892 &1503 & 1.54 \\
416 & 00 08 51 & 37 12 56 & 8.91 & A5   & 0.194 &  0.211 & 0.746 & 2.803 &-0.008 & -0.040  &2.064 &2174 & 2.30 \\
573 & 00 10 13 & 25 31 36 & 8.90& F0III & 0.208 & 0.253 & 0.783 & 2.780 &-0.057 & 0.043  &3.497  &1832 & 0.96\\
 & & & & & & & & & & & &2174 & 1.31\\
1607 & 00 20 27 & 22 28 47 & 8.67 & F0 & 0.254 & 0.222& 0.638  & 2.723 &-0.044    & 0.029 & 2.544 &2235 & 1.70 \\ 
2123 & 00 26 02 & 67 50 46 & 9.60 & F5 & -- & --      & -- & -- &-- &-- & -- &2223 & 1.30\\
2471 & 00 28 40 & 37 18 15 & 8.15 & A5 & 0.084 & 0.251 & 0.900 & 2.892 &-0.075  & -0.058  &2.449 &1499 & 2.05 \\ 
2523 & 00 29 08 & 11 19 12 & 8.08 & F0 & 0.219 & 0.208 & 0.655 &2.733  & -0.028 & 0.016 & 1.675 &2215 & 2.06 \\
 & & & & & & & & & & & &2223 & 1.97\\
 & & & & & & & & & & & &2283 & 0.79\\
4564 & 00 48 00 & 36 12 17 & 8.05 & A5 & 0.150 &  0.228 & 0.802 & 2.824&-0.022  & -0.024  & 2.290 &2214 & 2.12\\
7901 & 01 20 36 & 67 14 02 & 8.44 & A3 & 0.212 & 0.209 & 0.720 & 2.750 &-0.024  & 0.040  & 1.670 & 2240&1.05 \\
9550 & 01 34 24 & 38 50 44 & 8.29 & A3 & 0.199 & 0.213 & 0.740 & 2.762&-0.024  & 0.036  & 1.765 & 1827&1.14 \\
9843 & 01 36 41 & 35 36 59 & 8.22 & F8 & 0.286 & 0.179 & 0.735 & 2.706&-0.006 & 0.187  & 0.908 & 1917& 1.65 \\
& & & & & & & & & & && 1919 & 1.37\\
& & & & & & & & & & && 1923 & 1.44\\
& & & & & & & & & & & & \\
\hline
\end{tabular}
\end{center}
\end{table*}

\newpage

\newpage

\end{document}